\documentclass[12pt,a4paper,twoside]{article}
\usepackage[english]{babel}
\usepackage[T1]{fontenc}
\usepackage[utf8]{inputenc}
\usepackage{graphicx}
\usepackage{geometry}
\usepackage{amsmath}
\usepackage{amssymb} 
\usepackage{xcolor}
\usepackage{natbib}
\usepackage[legalpaper,bookmarks=true,colorlinks=true,linkcolor=blue,citecolor=blue]{hyperref}
\usepackage{color}
\usepackage{hyperref}
\usepackage[belowskip=-15pt,aboveskip=0pt]{caption}
\usepackage{subcaption}
\usepackage[toc,page]{appendix}
\usepackage{color}
\usepackage{color}
\usepackage{xcolor}
\usepackage{lettrine} 
\usepackage{latexsym}
\usepackage{eurosym}
\usepackage{lmodern}
\usepackage{fancyhdr}
\usepackage{float}
\usepackage{latexsym}
\begin{document}
\pdfoutput=1
\pagenumbering{arabic}
\begin{center}
\textbf{Bayesian estimation of the functional spatial lag model}
\end{center}
Alassane Aw, Laboratory of Mathematics and Applications, Assane Seck University of Ziguinchor, Ziguinchor, Senegal.\\
Email: alassane.aw.ansd@gmail.com.\\
\\
Emmanuel Nicolas Cabral, Laboratory of Mathematics and Applications, Assane Seck University of Ziguinchor, Ziguinchor, Senegal.\\
Email: encabral@univ-zig.sn.
\newpage
\section*{Abstract}
The spatial lag model (SLM) has been widely studied in the literature for spatialised data modeling in various disciplines such as geography, economics, demography, regional sciences, etc. This is an extension of the classical linear model that takes into account the proximity of spatial units in modeling. The extension of the SLM model in the functional framework (the FSLM model) as well as its estimation by the truncated maximum likelihood technique have been proposed by \cite{Ahmed}. In this paper, we propose a Bayesian estimation of the FSLM model. The Bayesian MCMC technique is used as estimation methods of the parameters of the model. A simulation study is conducted in order to compare the results of the Bayesian estimation method with the truncated maximum likelihood method. As an illustration, the proposed Bayesian method is used to establish a relationship between the unemployment rate and the curves of illiteracy rate  observed in the 45 departments of Senegal.
\\
\\
\noindent \textit{Keywords}. Spatial lag model, Functional data analysis, Bayesian estimation, MCMC algorithms.\\
\\
\noindent {\bfseries Mathematics Subject Classification :} 62F15, 62H11, 62P20, 60J22, 62-07.\\
\section{Introduction}
In the last decades, functional data is becoming increasingly common in many scientific fields, facilitated mainly by recent advances in data storage and processing. We can cite, among others, biology, climatology, econometrics, chemistry, meteorology that are likely to produce data considered as random curves. These data can be longitudinal, multivariate or spatial. Longitudinal data can include, for example, temperature curves, precipitation curves, growth curves of a plant, electrocardiogram of a patient, etc. Multivariate or logitidunal data may concern, for example, image modeling as random functions depending on two variables (abscissa and ordinate). The equipments for collecting and storing the functional data as powerful as they are, in fact, never makes it possible to observe a random curve. The functional data are therefore in vector form: they consist of a certain number of discrete values which have been measured on a sufficiently fine grid and recorded. Functional data analysis (FDA)  has arisen as a new discipline in the field of statistics proposing to model this type of data (\cite{Ramsay2005}).\\
The first category of models of the functional statistics are based on the hypothesis of the independence of the functions $X_{1}(t), X_{2}(t), \dots, X_{n}(t)$ defined on a compact interval $I=[0, T]$ on the real line $\mathbb{R}$. For instance, functional linear models (\cite{Cardot1999}, \cite{Cardot2003}), \cite{Hall}, \cite{Hilgert}), functional analysis of variance (\cite{Zoglat}, \cite{Zhang2013}), multivariate functional data (\cite{James}, \cite{Jacques}) are models that rely on the assumption of independence of functions. However, in many disciplines of applied sciences, there is a growing need to model corelated functional data. This is the case when samples of functions are observed over a discrete set of time points (temporally correlated functional data) or when these functions are observed over different sites or areas of a region (spatially correlated functional data). In such cases, the independence assumption of $X_{1}(t), X_{2}(t), \dots, X_{n}(t)$ is violated and the above models mentionned are not appropriate. To overcome these limitations, appropriate models are developed to take into account the dependency between functions.  For more developments, one can refer to the works of \cite{Delicado}, \cite{Ramsay2011}.\\
In the domain of spatially correlated data, functional geostatistics models random functions observed at a given set of points $s_{1}, s_{2}, \dots, s_{n}$ in a region $\mathcal{D}\subset\mathbb{R}^{n}$ (\cite{Dabo}, \cite{Giraldo}, \cite{Caballero}). Functional point processes correspond to the case where the functions are observed at each point $s_{1}, s_{2}, \dots, s_{n}$ in $\mathcal{D}$ which is generated by a standard point process. When $\mathcal{D}$ is a fixed and countable set, functional areal data (functional data in lattices) are concerned. Some works relating to functional areal data exist in the literature and can be found in \cite{Zhang2016}, \cite{Ahmed}, \cite{Huang}, \cite{Pineda}.\\
The motivation of this paper is to provide a Bayesian estimation method to functional linear models on lattices. Specifically, we are intersted at the Bayesian estimation of the spatial lag model which is  one of the family of spatial autoregressive models. This paper is organized as follows: we present the functional spatial lag model (FSLM) in the Bayesian context in section 2. In section 3, we propose the Bayesian MCMC estimation procedure of the functional spatial lag model. Section 4 gives a simulation study of the model proposed. In section 5, we apply the methodology proposed to real data. We end the article by a conclusion in section 6.
\section{The model}
We consider here n spatial units located in a fixed and countable region $\mathcal{D}\subset\mathbb{R}^{n}$. In each spatial unit, we observe a real response variable $y_{i}$  and a functional explanatory variable $X_{i}(t), t\in\mathcal{T}$ which takes its values in the Hilbert space $L^{2}(\mathcal{T})$. The set $\mathcal{T}$ is a compact interval of the real line $\mathbb{R}$. \\
We start from the model (see \cite{Ahmed}) which assumes an endogenous relationship between $y_{i}$ and $X_{i}(t)$ according to the spatial lag model defined by
\begin{equation}
\label{prop1}
y_{i}=\rho\sum_{j=1}^{n}\omega_{ij}y_{j}+\int_{\mathcal{T}}X_{i}(t)\gamma(t)dt+\epsilon_{i},\quad \quad \epsilon_{i}\sim N(0, \sigma^2), \quad i=1,2,\dots, n,
\end{equation}
where $\rho$ is an unkwon real spatial autoregressive parameter, $\gamma(t)$ is an unknown functional parameter, the error terms  $\epsilon_{i}$ are assumed to be independent and identically distributed according to the normal density with mean 0 and variance $\sigma^2$ and $\omega_{ij}$ is the coefficient of the spatial weight matrix W defined by $W=(\omega_{ij})_{1\leq i,j\leq n}$ where $\omega_{ij}=1$ if the areas $i$ and $j$ are contiguous and $\omega_{ij}=o$ if not. By convention, $\omega_{ii}=0$ for all $i=1,2,\dots, n$. \\
The introduction of this matrix is very important because it defines the relationships among locations. The weight matrix W is the space version of lag operator in time series \cite{Anselin}. In practice, the weight matrix W is row-standardized so that comparisons between models are easy to make.\\
Consider an orthonormal basis $\{\phi_{j}, j\in \mathbb{N}\}$ of $L^{2}(\mathcal{T})$. We can decompose $X_{i}(t)$ and $\gamma(t)$ in this basis as follows
\begin{equation}
X_{i}(t)=\sum_{j=1}^{\infty}z_{ij}\phi_{j}(t) \quad\mbox{and}\quad \gamma(t)=\sum_{j=1}^{\infty}\beta_{j}\phi_{j}(t).
\end{equation}
The real random variables $z_{ij}$ and the functional coefficients $\beta_{j}$ are given by
\begin{equation}
z_{ij}=\int_{\mathcal{T}}X_{i}(t)\phi_{j}(t)dt\quad\mbox{and}\quad\beta_{j}=\int_{\mathcal{T}}\gamma(t)\phi_{j}(t)dt.
\end{equation}
From this decomposition, it follows that
\begin{equation}
\int_{\mathcal{T}}X_{i}(t)\gamma(t)dt=\sum_{j=1}^{\infty}z_{ij}\beta_{j}.
\end{equation}
Equation (4) can be proved using dominated convergence and the fact that $\{\phi_{j}, j\in \mathbb{N}\}$ is an orthornormal basis of $L^{2}(\mathcal{T})$ by writing
\begin{eqnarray}
\int_{\mathcal{T}}X_{i}(t)\gamma(t)dt &=&\int_{\mathcal{T}}\left(\sum_{k=1}^{\infty}z_{ik}\phi_{k}(t)\right)\left(\sum_{j=1}^{\infty}\beta_{j}\phi_{j}(t)\right)dt\nonumber\\
                                     &=&\sum_{k=1}^{\infty}\sum_{j=1}^{\infty}z_{ik}\beta_{j}\left(\int_{\mathcal{T}}\phi_{k}(t)\phi_{j}(t))dt\right)\nonumber\\
                                     &=&\sum_{k=1}^{\infty}\sum_{j=1}^{\infty}z_{ik}\beta_{j}\langle\phi_{k}, \phi_{j} \rangle\nonumber\\
                                     &=&\sum_{k=1}^{\infty}\sum_{j=1}^{\infty}z_{ik}\beta_{j}\delta_{kj}\nonumber\\
                                     &=&\sum_{j=1}^{\infty}z_{ij}\beta_{j}\nonumber,
\end{eqnarray}
where $\langle\phi_{k}, \phi_{j} \rangle=\int_{\mathcal{T}}\phi_{k}(t)\phi_{j}(t))dt$ is the inner product of $L^{2}(\mathcal{T})$ of the functions $\phi_{k}, \phi_{j}$ and $\delta_{kj}$ is the  Kronecker symbol. \\
Following  \cite{Ahmed}, we decompose the right-hand side of equation (4) like this
\begin{equation}
\sum_{j=1}^{\infty}z_{ij}\beta_{j}=\sum_{j=1}^{k_{n}}z_{ij}\beta_{j}+\sum_{j=k_{n}+1}^{\infty}z_{ij}\beta_{j},
\end{equation}
where $k_{n}$ is a sequence of positive integers which increases with the size of the sample n. The goal here is to approach the infinite sum $\sum_{j=1}^{\infty}z_{ij}\beta_{j}$ by the finite sum $\sum_{j=1}^{k_{n}}z_{ij}\beta_{j}$. This is only possible when the sum $\sum_{j=k_{n}+1}^{\infty}z_{ij}\beta_{j}$ becomes negligible. For more details, one can refer to \cite{Muller} and \cite{Ahmed}.\\
The truncation of equation (5) at order $k_{n}$ can be accomplished by using the eingenfunctions basis, the Fourier basis, the spline basis or the wavelet basis. When the variable to be truncated is periodic, one can use the Fourier basis. In the other cases, one can think of using the other forms of basis according to the situations.\\
After truncation, model (1) becomes:
\begin{equation}
\label{prop1}
y_{i}=\rho\sum_{j=1}^{n}\omega_{ij}y_{j}+\sum_{j=1}^{k_{n}}z_{ij}\beta_{j}+\epsilon_{i},\quad \epsilon_{i}\sim N(0, \sigma^2), \quad i=1,2,\dots, n.
\end{equation}
Then, the infinite dimension of model (1) is reduced to a finite dimension. The problem now comes down to estimating $k_{n}+2$ parameters: $k_{n}$ parameters $\beta_{j}$ and 2 parameters concerning the spatial autoregressive parameter $\rho$ and the variance $\sigma^2$ of the  error term.\\
\\
If we set $y=\begin{bmatrix}
y_1\\
y_2\\
\vdots\\
y_n
\end{bmatrix}$, $\textbf{Z}_{k_{n}}=\begin{bmatrix}
    z_{11} & z_{12} & \dots  & z_{1k_{n}} \\
    z_{21} & z_{22} & \dots  & z_{2k_{n}} \\
    \vdots      & \vdots      & \ddots & \vdots \\
    z_{n1} & z_{n2} & \dots  & z_{nk_{n}}
\end{bmatrix}$, $\mathbf{\beta_{k_{n}}}=\begin{bmatrix}
\beta_{1}\\
\beta_{2}\\
\vdots\\
\beta_{k_{n}}\end{bmatrix}$, $\epsilon=\begin{bmatrix}
\epsilon_{1}\\
\epsilon_{2}\\
\vdots\\
\epsilon_{n}
\end{bmatrix}$, $W=\begin{bmatrix}
    \omega_{11} & \omega_{12} & \dots  & \omega_{1n} \\
    \omega_{21} & \omega_{22} & \dots  & \omega_{2n} \\
    \vdots      & \vdots      & \ddots & \vdots \\
    \omega_{n1} & \omega_{n2} & \dots  & \omega_{nn}
\end{bmatrix}$, \\
\\ then the matrix form of equation (6) can be written as
\begin{equation}
y=\rho Wy+\textbf{Z}_{k_{n}}\mathbf{\beta_{k_{n}}}+\epsilon, \quad \epsilon\sim N(0, \sigma^2I_{n}),
\end{equation}
where $I_{n}$ is the identity matrix of order n. \\
As in \cite{LeSage1997} and \cite{LeSage2009}, we assume a normal prior for $\mathbf{\beta_{k_{n}}}$ with mean $m_{k_{n}}$ and variance-covariance matrix $\Sigma_{k_{n}}$, an inverse gamma prior for $\sigma^2$ with shape parameter $a$ and scale parameter $b$ and a uniform prior for $\rho$. These prior distributions are given by
\begin{itemize}
\item $p(\mathbf{\beta_{k_{n}}})\sim N(m_{k_{n}}, \Sigma_{k_{n}})=\frac{1}{(2\pi)^{\frac{k_{n}}{2}}|\Sigma_{k_{n}}|^{\frac{1}{2}}}\exp\left(-\frac{1}{2}(\mathbf{\beta_{k_{n}}}-m_{k_{n}})^{T}\Sigma_{k_{n}}^{-1}(\mathbf{\beta_{k_{n}}}-m_{k_{n}})\right)$, \\
\item $p(\sigma^2)\sim IG(a, b)=\frac{b^a}{\Gamma(a)}(\sigma^2)^{-(a+1)}\exp\left(-\frac{b}{\sigma^2}\right)$,\\ 
\item $p(\rho)\sim U[0, 1]$.
\end{itemize}
In order to facilitate the statistical processing, we assume that all prior distributions are independent.
\section{Bayesian MCMC estimation of the FSLM}
\subsection{Full conditional distributions}
\lettrine From equation (7), we can deduce
\begin{eqnarray*}
\epsilon &=& y-\rho Wy-\textbf{Z}_{k_{n}}\mathbf{\beta_{k_{n}}}\\
         &=& (I_{n}-\rho W)y-\textbf{Z}_{k_{n}}\mathbf{\beta_{k_{n}}}\\
         &=& Ay-\textbf{Z}_{k_{n}}\mathbf{\beta_{k_{n}}}.
\end{eqnarray*}
where $A=I_{n}-\rho W$. Noting $\theta=(\mathbf{\beta_{k_{n}}}^T, \sigma^2,\rho)^T$ and using the transformation theorem, the likelihood function of the model is given by
\begin{equation*}
L(\theta)=f(y|\textbf{Z}_{k_{n}};\theta)=f(\epsilon|\textbf{Z}_{k_{n}};\theta)\left|\frac{\partial \epsilon}{\partial y}\right|.
\end{equation*}
where $\frac{\partial \epsilon}{\partial y}$ is the Jacobian matrix. Since $\epsilon\sim N(0, \sigma^2 I_{n})$, its multidimensional probability density is given by
\begin{eqnarray}
f(\epsilon|\textbf{Z}_{k_{n}};\theta) &=& \frac{1}{(2\pi)^{\frac{n}{2}}|\sigma^2I_{n}|^{\frac{1}{2}}}\exp\left[-\frac{1}{2}\epsilon^T(\sigma^2 I_{n})^{-1}\epsilon\right]\nonumber\\
                     &=& \frac{1}{(2\pi)^{\frac{n}{2}}(\sigma^2)^{\frac{n}{2}}}\exp\left[-\frac{1}{2\sigma^2}\epsilon^T\epsilon\right]\nonumber\\
                     &=& (2\pi)^{-\frac{n}{2}}(\sigma^2)^{-\frac{n}{2}}\exp\left[-\frac{1}{2\sigma^2}(Ay-\textbf{Z}_{k_{n}}\mathbf{\beta_{k_{n}}})^T (Ay-\textbf{Z}_{k_{n}}\mathbf{\beta_{k_{n}}})\right].
\end{eqnarray}
The calculation of the determinant of the Jacobian matrix is as follows
\begin{eqnarray}
\left|\frac{\partial \epsilon}{\partial y}\right| &=& \left|\frac{\partial (Ay-\textbf{Z}_{k_{n}}\mathbf{\beta_{k_{n}}})}{\partial y}\right|\nonumber\\
                                                  &=& \left|\frac{\partial Ay}{\partial y}-\frac{\partial \textbf{Z}_{k_{n}}\mathbf{\beta_{k_{n}}}}{\partial y}\right|\nonumber\\
                      &=& \left|\frac{\partial Ay}{\partial y}\right|\nonumber\\
                     &=& \left|A \right|\nonumber \\
                     &=& \left|I_{n}-\rho W\right|.
\end{eqnarray}
Combining equation (8) and equation (9), the likelihood function of the model is then given by
\begin{equation}
L(\theta)=f(y|\textbf{Z}_{k_{n}};\theta)=(2\pi)^{-\frac{n}{2}}(\sigma^2)^{-\frac{n}{2}}\exp\left[-\frac{1}{2\sigma^2}(Ay-\textbf{Z}_{k_{n}}\mathbf{\beta_{k_{n}}})^T (Ay-\textbf{Z}_{k_{n}}\mathbf{\beta_{k_{n}}})\right]\left|A \right|.
\end{equation}
The posterior distribution of the model, under the assumption of independence of the prior distributions,  is
\begin{equation}
p(\mathbf{\beta_{k_{n}}}, \sigma^2, \rho|y, W)=L(\theta)\times p(\mathbf{\beta_{k_{n}}})\times p(\sigma^2)\times p(\rho).
\end{equation}
When we ignore the constant terms involved in the prior distributions of the parameters, the posterior distribution can be put in the form of a relation of proportionality as follows\\
\\
$p(\mathbf{\beta_{k_{n}}}, \sigma^2, \rho|y, W)\propto (\sigma^2)^{-\frac{n}{2}}\exp\left(-\frac{1}{2\sigma^2}(Ay-\textbf{Z}_{k_{n}}\mathbf{\beta_{k_{n}}})^T (Ay-\textbf{Z}_{k_{n}}\mathbf{\beta_{k_{n}}})\right)\left|A \right|\times (\sigma^2)^{-(a+1)}\exp\left(-\frac{b}{\sigma^2}\right)$ \\
$\times\exp\left(-\frac{1}{2}(\mathbf{\beta_{k_{n}}}-m_{k_{n}})^{T}\Sigma_{k_{n}}^{-1}(\mathbf{\beta_{k_{n}}}-m_{k_{n}})\right)$.\\
\\
From this posterior distribution, we will deduce the full conditional distributions of the parameters.
We start by looking for the conditional distribution of $\sigma^2$. By omitting the elements that do not include the $\sigma^2$ parameter in the posterior distribution, one can get the full conditional distribution for this parameter in terms of proportionality relation as descibed below
\begin{eqnarray}
p(\sigma^2|\mathbf{\beta_{k_{n}}}, \rho) &\propto& (\sigma^2)^{-\frac{n}{2}}\exp\left(-\frac{1}{2\sigma^2}(Ay-\textbf{Z}_{k_{n}}\mathbf{\beta_{k_{n}}})^T (Ay-\textbf{Z}_{k_{n}}\mathbf{\beta_{k_{n}}})\right)\times(\sigma^2)^{-(a+1)}\exp\left(-\frac{b}{\sigma^2}\right)\nonumber\\
                             &\propto& (\sigma^2)^{-\left(\frac{n}{2}+a+1\right)}\exp\left(-\frac{(Ay-\textbf{Z}_{k_{n}}\mathbf{\beta_{k_{n}}})^T (Ay-\textbf{Z}_{k_{n}}\mathbf{\beta_{k_{n}}})+2b}{2\sigma^2}\right).
\end{eqnarray}
We recognize in equation (12) an inverse gamma distribution for $\sigma^2|\mathbf{\beta_{k_{n}}}, \rho$ with shape parameter $\frac{n}{2}+a$ and scale parameter $\frac{(Ay-\textbf{Z}_{k_{n}}\mathbf{\beta_{k_{n}}})^T (Ay-\textbf{Z}_{k_{n}}\mathbf{\beta_{k_{n}}})+2b}{2}$, which we materialize by writing
\begin{equation}
\sigma^2|\mathbf{\beta_{k_{n}}}, \rho\sim IG\left(\frac{n}{2}+a, \frac{(Ay-\textbf{Z}_{k_{n}}\mathbf{\beta_{k_{n}}})^T (Ay-\textbf{Z}_{k_{n}}\mathbf{\beta_{k_{n}}})+2b}{2}\right).
\end{equation}
Similarly, starting from the posterior distribution and considering only terms that include the $\mathbf{\beta_{k_{n}}}$ parameter, we get the full conditional distribution given below
\begin{eqnarray}
p(\mathbf{\beta_{k_{n}}}|\sigma^2, \rho) &\propto& \exp\left(-\frac{1}{2\sigma^2}(Ay-\textbf{Z}_{k_{n}}\mathbf{\beta_{k_{n}}})^T (Ay-\textbf{Z}_{k_{n}}\mathbf{\beta_{k_{n}}})\right)\times\exp\left(-\frac{1}{2}(\mathbf{\beta_{k_{n}}}-m_{k_{n}})^{T}\Sigma_{k_{n}}^{-1}(\mathbf{\beta_{k_{n}}}-m_{k_{n}}\right)\nonumber\\
                             &\propto& \exp\left(-\frac{1}{2}\left((Ay-\textbf{Z}_{k_{n}}\mathbf{\beta_{k_{n}}})^T(\sigma^2 I_{n})^{-1}(Ay-\textbf{Z}_{k_{n}}\mathbf{\beta_{k_{n}}})+(\mathbf{\beta_{k_{n}}}-m_{k_{n}})^{T}\Sigma_{k_{n}}^{-1}(\mathbf{\beta_{k_{n}}}-m_{k_{n}})\right)\right)\nonumber.
\end{eqnarray}
In order to recognize the distributional form of $\mathbf{\beta_{k_{n}}}|\sigma^2, \rho$, we complete the square by expending out the terms $(Ay-\textbf{Z}_{k_{n}}\mathbf{\beta_{k_{n}}})^T(\sigma^2 I_{n})^{-1}(Ay-\textbf{Z}_{k_{n}}\mathbf{\beta_{k_{n}}})$ and $(\mathbf{\beta_{k_{n}}}-m_{k_{n}})^{T}\Sigma_{k_{n}}^{-1}(\mathbf{\beta_{k_{n}}}-m_{k_{n}})$ (we can refer to \cite{Gelman} for more details on the technique of completing the square).
We begin completing the square by wrting
\begin{equation*}
\left(Ay-\textbf{Z}_{k_{n}}\mathbf{\beta_{k_{n}}}\right)^T (\sigma^2 I_{n})^{-1}(Ay-\textbf{Z}_{k_{n}}\mathbf{\beta_{k_{n}}})=\mathbf{\beta_{k_{n}}}^T\left(\frac{1}{\sigma^2}\textbf{Z}_{k_{n}}^T\textbf{Z}_{k_{n}}\right)-\mathbf{\beta_{k_{n}}}^T\left(\frac{1}{\sigma^2}\textbf{Z}_{k_{n}}^T Ay\right)-\left(\frac{1}{\sigma^2}y^TA^T\textbf{Z}_{k_{n}}\right)\mathbf{\beta_{k_{n}}}+C_{1}.
\end{equation*}
where $C_{1}=y^TA^T\left(\sigma^2I_{n}\right)^{-1}Ay$. Similarly, we write
\begin{equation*}
(\mathbf{\beta_{k_{n}}}-m_{k_{n}})^{T}\Sigma_{k_{n}}^{-1}(\mathbf{\beta_{k_{n}}}-m_{k_{n}})=\mathbf{\beta_{k_{n}}}^T\Sigma_{k_{n}}^{-1}\mathbf{\beta_{k_{n}}}-\mathbf{\beta_{k_{n}}}^T\Sigma_{k_{n}}^{-1}m_{k_{n}}-m_{k_{n}}^T\Sigma_{k_{n}}^{-1}\mathbf{\beta_{k_{n}}}+m_{k_{n}}^T\Sigma_{k_{n}}^{-1}m_{k_{n}}.
\end{equation*}
After completing the square, the distribution of $\mathbf{\beta_{k_{n}}}|\sigma^2, \rho$ is a multinormal distribution with mean=$\left(\textbf{Z}_{k_{n}}^T\textbf{Z}_{k_{n}}+\sigma^2\Sigma_{k_{n}}^{-1}\right)^{-1}\left(\textbf{Z}_{k_{n}}^TAy+\sigma^2\Sigma_{k_{n}}^{-1}m_{k_{n}}\right)$ and variance-covariance matrix equals to $\left(\textbf{Z}_{k_{n}}^T\textbf{Z}_{k_{n}}+\sigma^2\Sigma_{k_{n}}^{-1}\right)^{-1}$. We summarize the conditional distribution of $\mathbf{\beta_{k_{n}}}|\sigma^2, \rho$ as follows
\begin{equation}
\mathbf{\beta_{k_{n}}}|\sigma^2, \rho\sim N\left(\left(\textbf{Z}_{k_{n}}^T\textbf{Z}_{k_{n}}+\sigma^2\Sigma_{k_{n}}^{-1}\right)^{-1}\left(\textbf{Z}_{k_{n}}^TAy+\sigma^2\Sigma_{k_{n}}^{-1}m_{k_{n}}\right), \left(\textbf{Z}_{k_{n}}^T\textbf{Z}_{k_{n}}+\sigma^2\Sigma_{k_{n}}^{-1}\right)^{-1}\right)
\end{equation}
The proportionality relation relative to the full conditional distribution for the parameter $\rho$ is given by
\begin{eqnarray}
p(\rho|\mathbf{\beta_{k_{n}}}, \sigma^2) &\propto& \left|A \right|\exp\left(-\frac{1}{2\sigma^2}(Ay-\textbf{Z}_{k_{n}}\mathbf{\beta_{k_{n}}})^T (Ay-\textbf{Z}_{k_{n}}\mathbf{\beta_{k_{n}}})\right)\nonumber\\
                             &\propto& \left|I_{n}-\rho W \right|\exp\left(-\frac{1}{2\sigma^2}\left((I_{n}-\rho W)y-\textbf{Z}_{k_{n}}\mathbf{\beta_{k_{n}}}\right)^T \left((I_{n}-\rho W)y-\textbf{Z}_{k_{n}}\mathbf{\beta_{k_{n}}}\right)\right)\nonumber.
\end{eqnarray}
Unfortunately there is any recognizable form for the full conditional distribution of the parameter $\rho$ due to the presence of the derminant $\left|I_{n}-\rho W \right|$.
\subsection{Bayesian MCMC computation}
Recall that $\theta=(\mathbf{\beta_{k_{n}}}^T, \sigma^2,\rho)^T$ is the parameter vector. We consider here the loss function $L(\hat{\theta}, \theta)=(\hat{\theta}-\theta)R(\hat{\theta}-\theta)$ where R is a positive define matrix. This function is used as a good criterion to determine an optimal point estimate of the parameter $\theta$. The Bayesian point estimate of the parameter $\theta$ is then given by
\begin{equation}
\hat{\theta}=argmin\mathbb{E}_{\theta}\left(L(\hat{\theta}, \theta)\right)=argmin\int_{\Theta}(\hat{\theta}-\theta)R(\hat{\theta}-\theta)p(\theta|y)d\theta.
\end{equation}
The solution of equation (15) is defined as $\hat{\theta}=\mathbb{E}(\theta|y)=\int_{\Theta}\theta p(\theta|y)d\theta$, which corresponds to the mean of the posterior distribution of $\theta$ (\cite{Koop}). In general, the analytical solution of equation (15) cannot be obtained in the closed form.\\
To avoid this problem, we use MCMC algorithms which consist of onstructing a Markov chain from the conditional posterior distributions. If the Markov chain is simulated long enough, then the mean of the sample is seen as an estimate of $\mathbb{E}(\theta|y)$ (\cite{Casella}, \cite{Roberts}).\\
Since we know the full conditional distributions for $\mathbf{\beta_{k_{n}}}$ and $\sigma^2$, we can use Gibbs sampling method to easily obtain random draws for these parameters. Unfortunately, the full conditional distribution of the autoregressive parameter $\rho$ is in the unknown form. Thus, we use the Metropolis-Hastings algorithm to obtain samples for this parameter. In practice, we combine the Gibbs sampling method and the Metropolis-Hastings algorithm for the sampling scheme of the functional spatial lag model. This technique is known in the litterature as the  Metropolis-within-Gibbs algorithm (\cite{Gilks}).\\
The Metropolis-Hastings algorithm uses a proposal distribution, according to which, we will sample the autoregressive parameter $\rho$. Following \cite{LeSage2009}, we can use a normal proposal distribution with tuned random-walk procedure defined as
\begin{equation}
\rho^{new}=\rho^{old}+c\psi, \quad \psi\sim N(0, 1).
\end{equation}
The constant c represents the tuning parameter. Since $ \psi$ is symmetric, so the probability to accept the candidate $\rho^{new}$ is given by
\begin{equation}
\alpha(\rho^{new}, \rho^{old})=\min\left(\frac{p(\rho^{new}|\mathbf{\beta_{k_{n}}}, \sigma^2, y)}{p(\rho^{old}|\mathbf{\beta_{k_{n}}}, \sigma^2, y)}, 1\right).
\end{equation}
\cite{Dogan} show that the tuning parameter affects the behavior of the chain in at least two ways: (i) it affects the acceptance rate of new candidate values through acceptance probability, (ii) it also affects the region where the new candidate values are sampled. \\
In practice, the tuning parameter is choosen so that the acceptance rate falls between 40$\%$ and 60$\%$.
The Metropolis-within-Gibbs algorithm for estimating parameters of the functional spatial lag model is described through the following steps.\\
1. Set up initial value $\theta^{(0)}=(\mathbf{\beta_{k_{n}}}^{(0)}, \sigma^{2(0)}, \rho^{(0)})$,\\
2. For iteration j from 1 to N, do the following steps.\\
2.1 Simulate $\mathbf{\beta_{k_{n}}}^{(j)}$ from $p(\mathbf{\beta_{k_{n}}}|\sigma^{2(j-1)}, \rho^{(j-1)})$,\\
2.2 Simulate $\sigma^{2(j)}$ from $p(\sigma^2|\mathbf{\beta_{k_{n}}}^{(j)}, \rho^{(j-1}))$,\\
2.3 Compute $\rho^{new}=\rho^{(j-1)}+c\psi$,\\
2.3.1 Calculate $\alpha(\rho^{new}, \rho^{(j-1)})=\min\left(\frac{p(\rho^{new}|\mathbf{\beta_{k_{n}}}^{(j)}, \sigma^{2(j)}, y)}{p(\rho^{(j-1)}|\mathbf{\beta_{k_{n}}}^{(j)}, \sigma^{2(j)}, y)}, 1\right)$,\\
2.3.2 Simulate u from uniform (0, 1),\\
2.3.3 Set $\rho^{(j)}=\rho^{new}$ if $\alpha(\rho^{new}, \rho^{(j-1)})>u$ else $\rho^{(j)}=\rho^{(j-1)}$.\\
2.3.4 Return $\theta^{(j)}=(\mathbf{\beta_{k_{n}}}^{(j)}, \sigma^{2(j)}, \rho^{(j)})$.
\\
The sequence $(\theta^{(0)}, \theta^{(1)}, \dots, \theta^{(N)})$ generated by the Metropolis-within-Gibbs algorithm is a markov chain whose stationary distribution converges towards the joint posterior density $p(\theta|y)$ (\cite{Thierney}).\\
Alternatively, we can use this algorithm considering a uniform proposal distribution instead of a normal proposal distribution. A slight modification of the algorithm takes place at the level of updating the autoregressive parameter $\rho$ (see, for instance  \cite{LeSage1997}, \cite{LeSage2009}).
\section{Simulation study} 
In this section, we propose simulations of the FSLM model from the point of view of the Bayesian approach. The simulations cover the 121 communes of Senegal represented by the spatial layout in figure 1 below.\\
\begin{figure}[H]
\caption{Spatial layout of communes of Senegal (left panel) and centroids of this communes (right
panel).}  
\centering
\includegraphics[width=16cm,height=8cm]{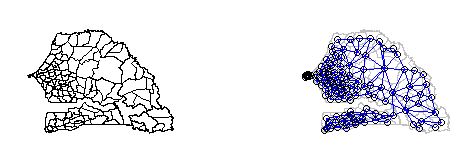}
\end{figure}
\noindent We consider several scenarios of simulations according to the values of the autoregressive parameter $\rho$. In each scenario, we compare the results obtained for the Bayesian FSLM model with uniform kernel and normal kernel, and the FSLM with maximum likelihood procedure. The process is summarized by the following steps.
\begin{itemize}
\item The $121\times 121$ weighting matrix W whose elements $w_{ij}$ are equal to 1 if the communes i and j share a common border and 0 otherwise is calculated through the spatial layout in figure 1;
\item In each commune, we simulate the functional covariate $X_{i}(t_{j})=cos(t_{j})+sin(t_{j})+\epsilon_{i}(t_{j}), \quad i = 1,2,\dots, 121, \quad j = 0,1, \dots, 100$, where $\epsilon_{i}(t_{j})$ is a realization of a white noise. We then smooth these values using a 7 B-spline basis in order to obtain the curves $X_{i}(t), \quad i=1,2,\dots, 121$. The number of basis is choosen using cross validation criterion (one can refer to Ramsay and Silverman (2005) for more details);
\item The functional parameter is defined by $\gamma(t)=e^{-\frac{t}{10}}\left(\left(\frac{t}{10}\right)^2+3\left(\frac{t}{10}\right)-4\right)$ as in \cite{Pineda};
\item We generate a $121\times 1$ Gaussian vector $\epsilon\sim N(0, \sigma^2 I_{n})$;
\item We calculate $y=(I_{n}-\rho W)^{-1}\textbf{Z}_{k_{n}}\mathbf{\beta_{k_{n}}}+(I_{n}-\rho W)^{-1}\epsilon$ by considering the values $\rho=0.3, 0.5, 0.7$;
\item The parameters $\gamma(t)$, $\sigma^2$ and $\rho$ are estimated using the Metropolis-within-Gibbs algorithm described above. We consider both the uniform kernel and the normal kernel for the proposal distribution.
\end{itemize}
The curves $X_{i}(t)$ are represented in figure 2 below.\\
\begin{figure}[H]
\caption{121 curves $X_{i}(t)$ obtained after smoothing the values $X_{i}(t_{j})$ by a 7 B-spline basis.}  
\centering
\includegraphics[width=16cm,height=8cm]{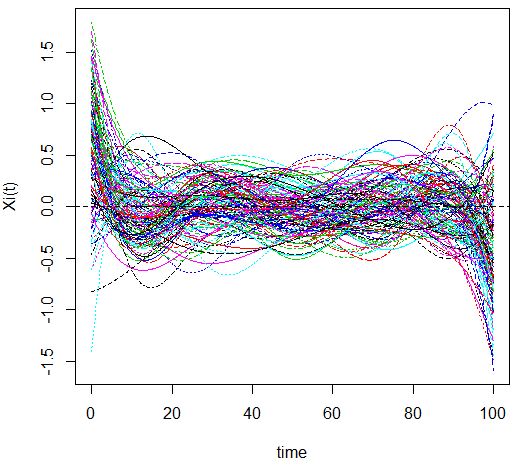}
\end{figure}
\noindent We give in Table 1 below the results of simulations of the Metropolis-within-Gibbs algorithm for the FSLM model. For $\rho=0.3$, we find that all the methods considered give substantially the same results for the values of the $\beta$ parameter. However, it is generally noted that the normal kernel method and the maximum likelihood method are closer in terms of the results obtained. The value of the simulated $\rho$ parameter ($\rho=0.28)$ for the normal kernel method is closer to the true value $\rho=0.3$ than the other methods. However, the uniform kernel method is preferable than the other methods according to the BIC values. The simulated values of the parameters are of the same order of magnitude for the different methods considered when $\rho=0.5$. 
\begin{table}[!htbp]
\centering
\caption{Comparison of simulation results of the Bayesian FSLM model}
\vspace{0.5cm}
\begin{tabular}{p{1.0cm} l l c c c c c c c c c}
\hline
$\rho$& Method &$\beta_{1}$& $\beta_{2}$ & $\beta_{3}$  &$\beta_{4}$  &$\beta_{5}$ & $\beta_{6}$&$\beta_{7}$ & $\sigma^2$& $\rho$& BIC\\
 \hline
 &Uniform kernel&-3.62 &-3.04 &-2.55 &-1.96 &-1.47& -1.64&-0.79&1.09&0.48&182 \\

0.3 & Normal kernel&-3.67& -3.06 &-2.53 & -1.98& -1.46& -1.15&-0.82&1.07&0.28&185 \\
 & Maximum likelihood& -3.68& -3.05&-2.52 & -1.90& -1.43& -1.16&-0.80&1.00&0.25&393 \\
\hline
 &Uniform kernel &-3.49 &-2.82 &-2.26 &-1.41 &-1.27& -0.86&-0.35&0.99&0.54&188 \\

0.5 &Normal kernel& -3.41& -2.76&-2.20 & -1.33& -1.23& -0.86&-0.40&0.97&0.46&203 \\
 & Maximum likelihood& -3.45& -2.80 &-2.26 & -1.41& -1.30& -0.88&-0.39&0.89&0.44&383 \\
\hline
  &Uniform kernel&-3.32 &-2.76 &-2.18 &-1.32 &-1.20& -0.91&-0.51&0.81&0.57&427 \\

0.7 &Normal kernel& -3.28& -2.73&-2.20 & -1.37& -1.23& -0.91&-0.54&0.81&0.72&289 \\
 & Maximum likelihood & -3.33& -2.75&-2.18 & -1.33& -1.22& -0.89&-0.50&0.74&0.74&371 \\
\hline
\end{tabular}
\end{table}
\noindent However, the uniform kernel and maximum likelihood methods produce simulated values that are substantially the same. It is always noted that the uniform kernel method is preferable with respect to the value of the BIC.\\
When $\rho=0.7$, it was again found that the uniform kernel and maximum likelihood methods yield results that are about the same. The normal kernel method gives a value $\rho=0.72$ which is closer to the true value $\rho=0.7$. \\This method, with a value of $BIC=289$, is preferable. In summary, the general finding is that the Bayesian proposition is better than the classical method. When the $\rho$ value is low, the uniform kernel method overrides the normal kernel method. However for large $\rho$ values, the normal kernel method is preferable to the uniform kernel method.

\section{Application to unemployment data}
In this section, we apply the proposed methodology to real data. The aim is to explain the variations in the unemployment rate according to those of the illiteracy rate. In each department of Senegal, we observe the unemployment rate in the first quarter of 2019 and the illiteracy rate from the second quarter of 2016 to the first quarter of 2019. The data come from the National Agency of Statistics and Demography of Senegal.\\
We give in Figure 3 below a spatial distribution of the unemployment rate in each of the 45 departments of Senegal. We observe in this figure that the spatial distribution of the unemployment rate is not neutral. Most geographically close departments have similar unemployment values. \\

\begin{figure}[H]
\caption{Spatial distribution of the unemployment rate in Senegal.} 
\centering
\includegraphics[width=16cm,height=8cm]{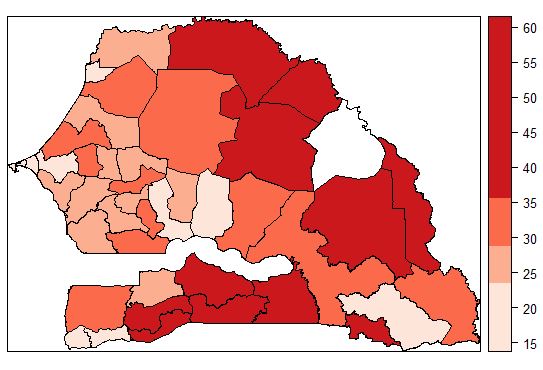}
\end{figure}
\noindent This means that the unemployment rate is a spatially autocorrelated variable. This visual observation was confirmed by the Moran test.\\
Figure 4 below shows the curves of the illiteracy rate over the observation period. As a first step, these discrete data on the illiteracy rate collected per quarter are smoothed using 7 B-splines basis.
Overall, the curves of the illiteracy rate are decreasing over the observation period. This shows the great efforts made by the Government of Senegal to reduce the illiteracy rate by formulating targeted literacy programs.\\
\begin{figure}[H]
\caption{Curves of the rate of illiteracy of Senegal, obtained after expanding the data by using a 7 B-splines basis.} 
\centering
\includegraphics[width=16cm,height=8cm]{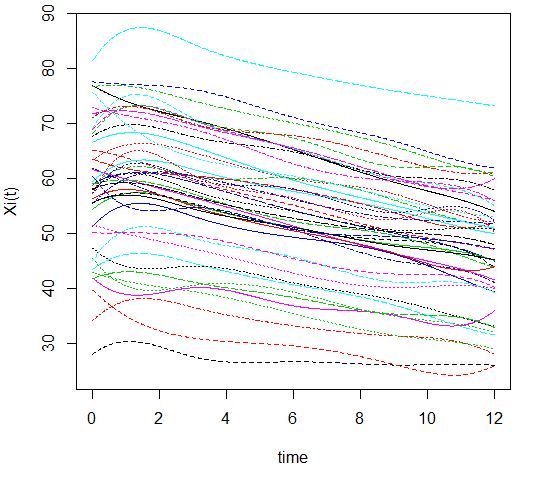}
\end{figure}

 


\noindent In Table 2 below, we group the results of the parameter estimation of the Bayesian FSLM model according to the different methods. The results of the estimation of the vector parameter $\beta$ for the normal kernel and the uniform kernel methods that are very close do not deviate too much from the results obtained for the maximum likelihood method. The estimated standard deviations of the normal kernel and the uniform kernel methods are almost the same. The values of the $\rho$ parameter for the normal kernel and the uniform kernel methods show that the autocorrelation of the unemployment rate is more pronounced in these models than that of the maximum likelihood.\\

 

\begin{table}[!htbp]
\centering
\caption{Comparison of parameter estimation of the Bayesian FSLM model}
\vspace{0.5cm}
\begin{tabular}{p{3cm} p{1cm} l c c c c c c c c c c}
\hline
Method&  &$\beta_{1}$& $\beta_{2}$ & $\beta_{3}$  &$\beta_{4}$  &$\beta_{5}$ & $\beta_{6}$&$\beta_{7}$ & $\sigma^2$& $\rho$& BIC & AR \\
 \hline
 &Value&-0.31 &-0.22 &1.08 &0.89 &-1.45& 0.26&-0.09& 57&0.53&299&69$\%$ \\
Normal kernel  \\
 & Std& 0.38& 0.47&0.67 & 0.84& 0.75& 0.69&0.77& 14& 0.14& \\
\hline
 &Value&-0.32 &-0.24 &1.21 &1.03 &-1.58& 0.17&-0.07&58&0.53&279&12$\%$ \\
Uniform kernel  \\
 & Std& 0.38& 0.48&0.68 & 0.86& 0.77& 0.70&0.79& 14&0.27& \\
 \hline
 &Value&-0.27 &-0.22 &1.23 &0.99 &-1.59& 0.20&-0.08&45&0.45&340 \\
ML method  \\
 & Std& 0.34& 0.42&0.60 & 0.74& 0.67& 0.61&0.69& 6.73&0.14& \\
 
\hline
\end{tabular}
\end{table}
\noindent Looking at the values of the BIC, we conclude that the uniform kernel method gives the best model. However, the normal kernel method is preferable if we look at the acceptance rate (AR=69$\%$) of the Metropolis-within-Gibbs algorithm. In all cases, the results of the Bayesian approach are better than those obtained by the frequentist approach.\\

\section{Conclusion}
In this paper, we have proposed a Bayesian estimation method of the functional spatial lag model. Using the Metropolis-within-Gibbs algorithm, we have shown on the basis of numerical simulations that Bayesian estimation gives better results than frequentist estimation. We have also illustrated the Bayesian methodology with an application to unemployement data of Senegal. The results obtained confirm that the Bayesian methodology is more satisfactory than the maximum likelihood estimation. This modeling
approach is innovative in the fields of spatial statistics and spatial econometrics. The application of this Bayesian methodology to other types of data and other areas of activity may be considered.

\begin{appendices}
The Markov chains for each parameter of the Metropolis-within-Gibbs algorithm are given in Figure 5 and figure 6 below. The chains of the different parameters of the Metropolis-within-Gibbs algorithm with normal kernel and uniform kernel are globally stable. The pace of the different chains also testify the speed of convergence of the algorithm. 
\begin{figure}
\caption{Markov chains produced by the Metropolis-within-Gibss algorithm with normal kernel.} 
  \begin{subfigure}[b]{0.3296293\textwidth}
    \includegraphics[width=\textwidth]{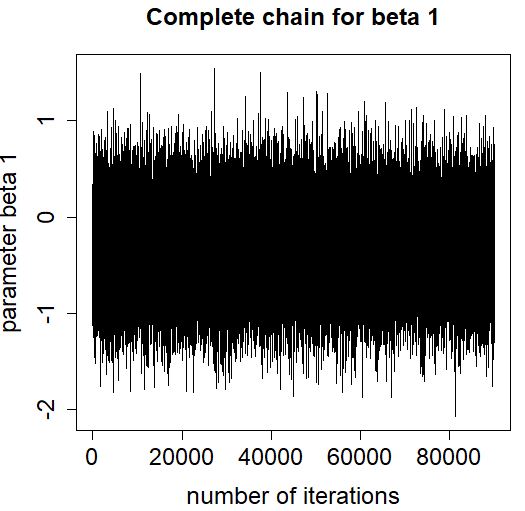}
    \label{fig:1}
  \end{subfigure}
  \begin{subfigure}[b]{0.3296293\textwidth}
    \includegraphics[width=\textwidth]{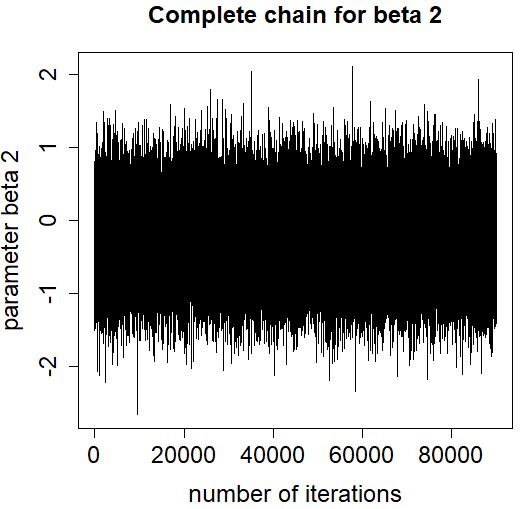}
    \label{fig:2}
    \end{subfigure}
  \begin{subfigure}[b]{0.3296293\textwidth}
    \includegraphics[width=\textwidth]{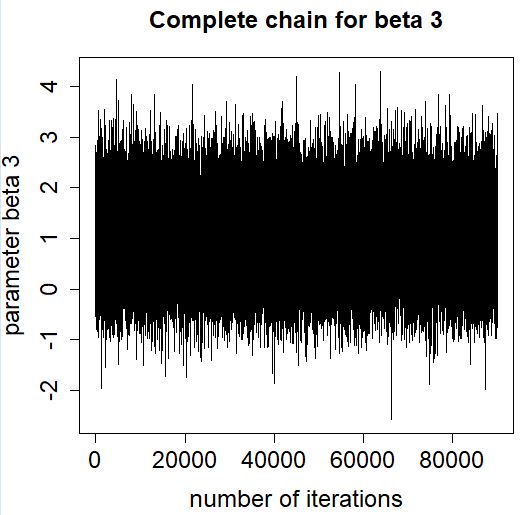}
    \end{subfigure}
  \begin{subfigure}[b]{0.3296293\textwidth}
    \includegraphics[width=\textwidth]{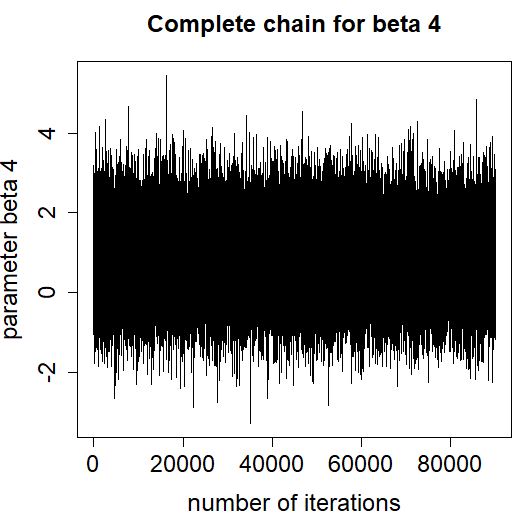}
    \end{subfigure}
  \begin{subfigure}[b]{0.3296293\textwidth}
    \includegraphics[width=\textwidth]{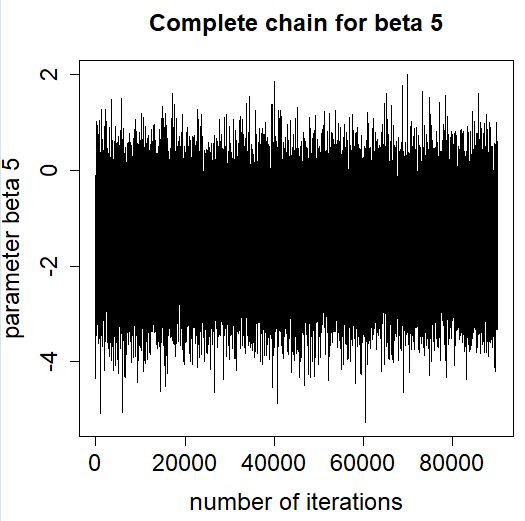}
    \end{subfigure}
  \begin{subfigure}[b]{0.3296293\textwidth}
    \includegraphics[width=\textwidth]{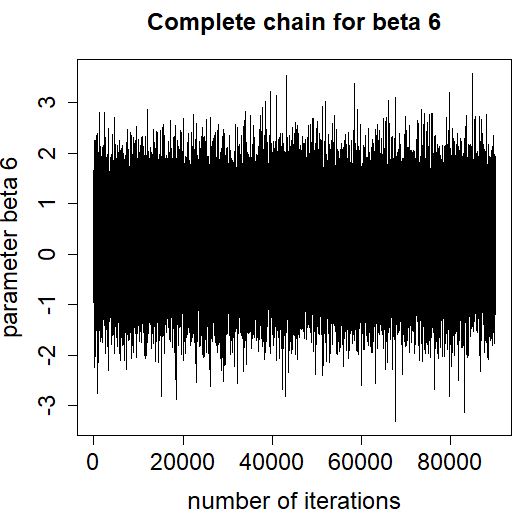}
    \end{subfigure}
  \begin{subfigure}[b]{0.3296293\textwidth}
    \includegraphics[width=\textwidth]{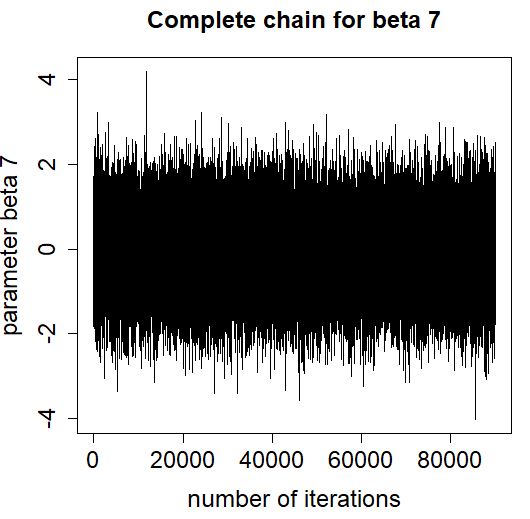}
    \end{subfigure}
  \begin{subfigure}[b]{0.32962593\textwidth}
    \includegraphics[width=\textwidth]{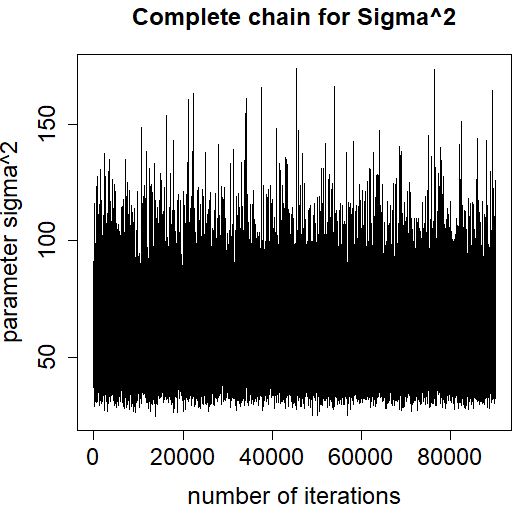}
    \end{subfigure}
  \begin{subfigure}[b]{0.32962593\textwidth}
    \includegraphics[width=\textwidth]{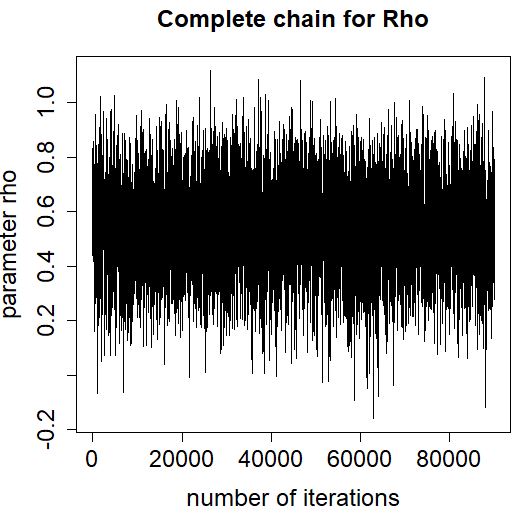}
    \end{subfigure}
\end{figure}
\begin{figure}
\caption{Markov chains produced by the Metropolis-within-Gibss algorithm with uniform kernel.} 
  \begin{subfigure}[b]{0.3296293\textwidth}
    \includegraphics[width=\textwidth]{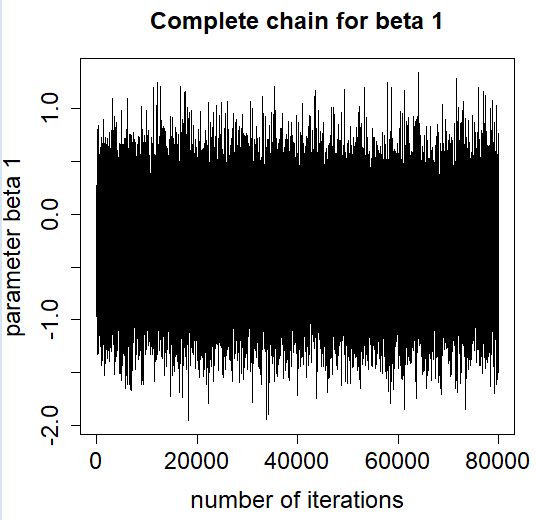}
    \label{fig:1}
  \end{subfigure}
  \begin{subfigure}[b]{0.3296293\textwidth}
    \includegraphics[width=\textwidth]{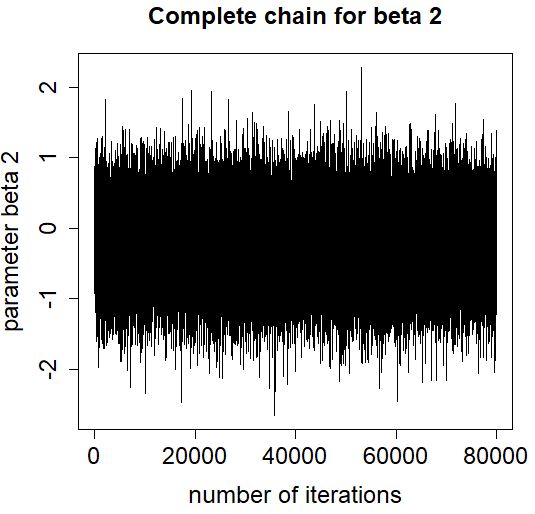}
    \label{fig:2}
    \end{subfigure}
  \begin{subfigure}[b]{0.3296293\textwidth}
    \includegraphics[width=\textwidth]{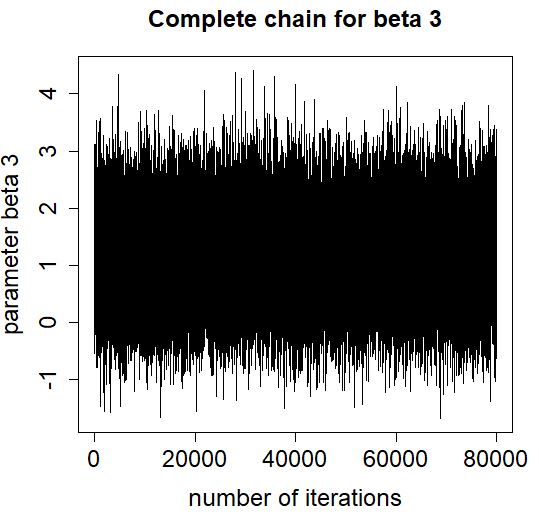}
    \end{subfigure}
  \begin{subfigure}[b]{0.3296293\textwidth}
    \includegraphics[width=\textwidth]{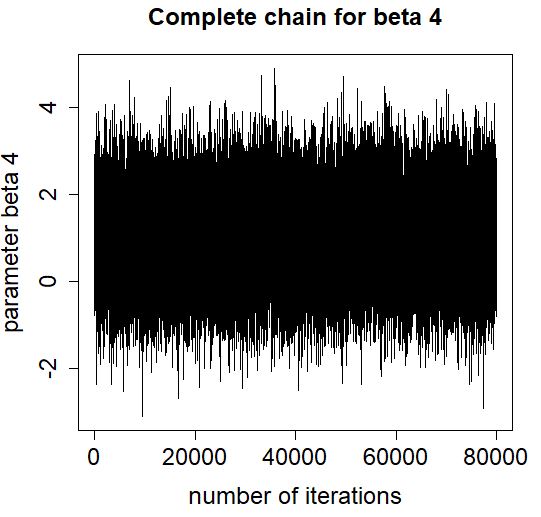}
    \end{subfigure}
  \begin{subfigure}[b]{0.3296293\textwidth}
    \includegraphics[width=\textwidth]{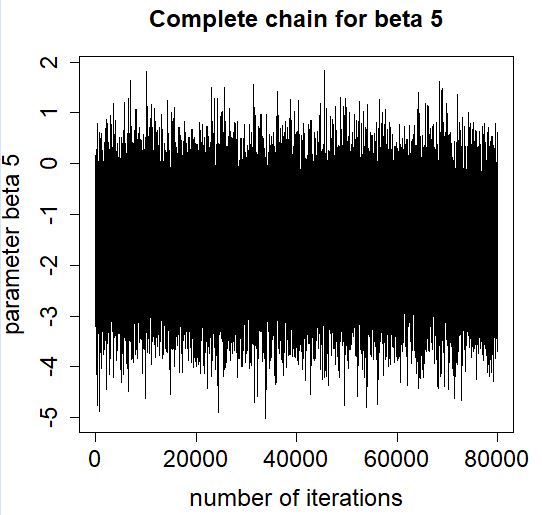}
    \end{subfigure}
  \begin{subfigure}[b]{0.3296293\textwidth}
    \includegraphics[width=\textwidth]{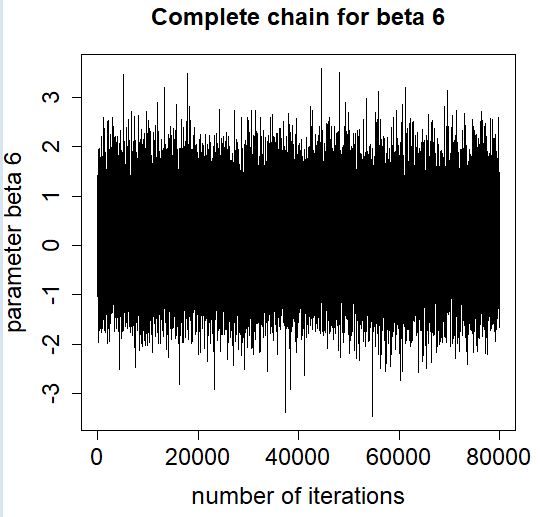}
    \end{subfigure}
  \begin{subfigure}[b]{0.3296293\textwidth}
    \includegraphics[width=\textwidth]{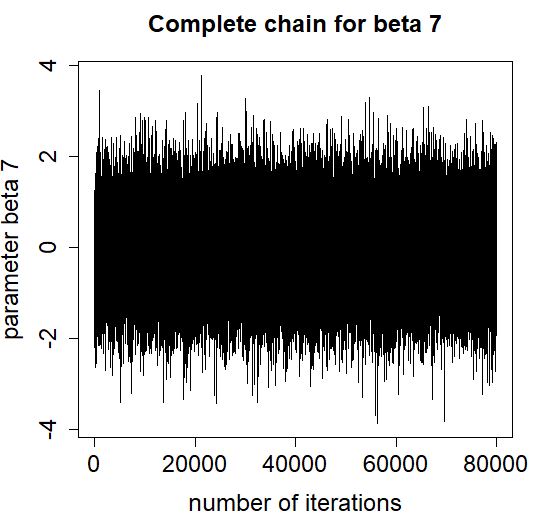}
    \end{subfigure}
  \begin{subfigure}[b]{0.32962593\textwidth}
    \includegraphics[width=\textwidth]{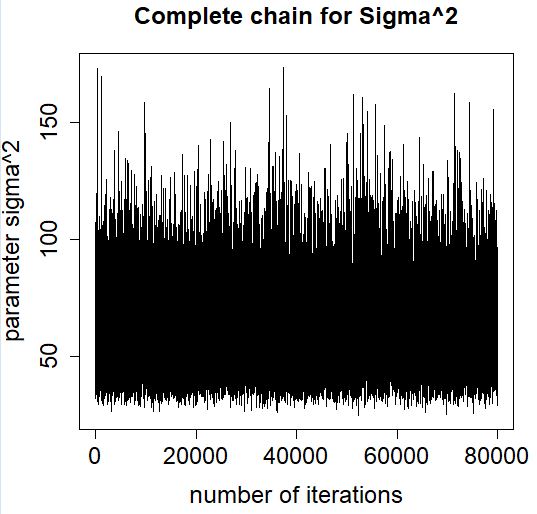}
    \end{subfigure}
  \begin{subfigure}[b]{0.32962593\textwidth}
    \includegraphics[width=\textwidth]{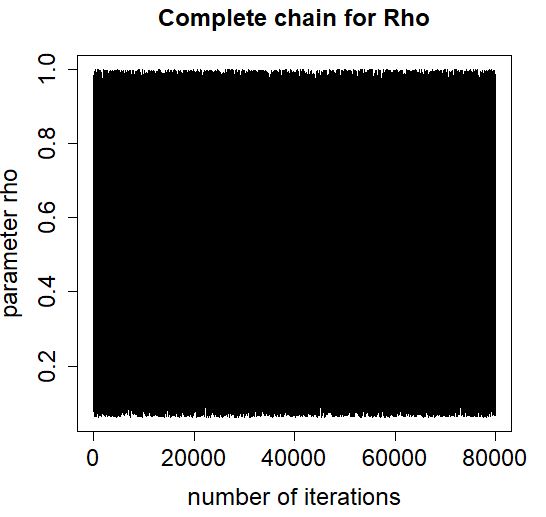}
    \end{subfigure}
\end{figure}
\end{appendices}
\end{document}